\documentclass[preprint,12pt]{elsarticle}

\usepackage{amssymb}
\usepackage{xcolor}
\usepackage{soul}
\usepackage{rotating}
\usepackage{makecell}
\usepackage{amsmath}

\journal{Machine Learning with Applications}

\begin{document}

\begin{frontmatter}


\title{A Natural Language Processing and Deep Learning based Model for Automated Vehicle Diagnostics using Free-Text Customer Service Reports}
Authors: Ali Khodadadi (1), Soroush Ghandiparsi (1), Chen-Nee Chuah (2)
       ((1) Department of Electrical and Computer Engineering,  University of California,545 Bainer Hall Dr,Davis,95616,CA,USA)


\begin{abstract}
Initial fault detection and diagnostics are imperative measures to improve the efﬁciency, safety, and stability of vehicle operation. In recent years, numerous studies have investigated data-driven approaches to improve the vehicle diagnostics process using available vehicle data. Moreover, data-driven methods are employed to enhance customer-service agent interactions. In this study, we demonstrate a machine learning pipeline to improve automated vehicle diagnostics. First, Natural Language Processing (NLP) is used to automate the extraction of crucial information from free-text failure reports (generated during customers' calls to the service department). Then, deep learning algorithms are employed to validate service requests and filter vague or misleading claims. Ultimately, different classification algorithms are implemented to classify service requests so that valid service requests can be directed to the relevant service department. The proposed model- Bidirectional Long Short Term Memory (BiLSTM) along with Convolution Neural Network (CNN)- shows more than 18\% accuracy improvement in validating service requests compared to technicians' capabilities. In addition, using domain-based NLP techniques at preprocessing and feature extraction stages along with CNN-BiLSTM based request validation enhanced the accuracy ($>$25\%), sensitivity ($>$39\%), specificity ($>$11\%), and precision ($>$11\%) of Gradient Tree Boosting (GTB) service classification model. The Receiver Operating Characteristic Area Under the Curve (ROC-AUC) reached 0.82.
\end{abstract}
\begin{keyword}
Customer Service\sep NLP\sep CNN\sep BiLSTM\sep claim validation\sep Classification\sep Vehicle Diagnostic, Content Extraction\sep Automotive Data Mining
\end{keyword}
\end{frontmatter}
\section{INTRODUCTION}
\label{intro}
Data science and ML techniques have led to groundbreaking changes in the automotive industry. By 2030, the automotive sector can upgrade its products with autonomous mobility solutions, vehicle-to-vehicle, and vehicle–to-network communication systems \cite{1}.  Efficient customer service is a critical factor for future transportation success. Failed vehicle early-stage diagnostics is crucial for research and development sections to determine product improvement strategies. In addition, efficient diagnostics leads to the more successful sale of both products and services.
In 2017, U.S. businesses lost 75 billion dollars due to inefficient and lack of timely customer service procedures that significantly raised the level of customer unsatisfactory. For instance, based on Microsoft state of global customer service reports, 22 \% of clients contacting the customer service section had to repeat their service request \cite{2}. Current vehicle service systems address vehicle failures via two main approaches \cite{3}:
\begin{itemize}
\item Use On-Board Diagnostic (OBD) systems via specific toolsets. The customer or technician needs to communicate with the Electronic Controller Unit (ECU) via a scanner tool. The ECU unit usually provides fault codes, or the operator can record continuous signals from the onboard vehicle sensors, actuators, and controllers during diagnosis operation.
\item The client initiates a service request to the service sector, describes the issue to the best of his knowledge, primarily in free text format, and pursues “on-the-road” services support if further assistance is required.
\end{itemize}
\subsection{Related Work}
There are numerous common limitations on the OBD systems. They are typically designed for specific vehicle manufacturers; Thus, they cannot be used across multiple vehicle companies. Even the OBD-II or J1939 standards are designed to address generic vehicle issues and are not capable of in-depth vehicle diagnostics. Moreover, due to vehicle hardware computation limitations, diagnostic logic is statics and not user-friendly. Therefore, customers typically prefer to contact the service section to solve the problem. However, extending the “human-based customer service” is not desirable for companies as each case imposes extra charges, including labor, office, and on-the-road support expenses.

Improving vehicle customer service will be crucial for the next generation of intelligent transportation systems. In recent years, researchers have focused on data-driven fault detection algorithms to enhance diagnostic methods. However, most of the complex algorithms studied require a large amount of data to train and fit into an efficient model. Sufficient data for such models is a challenging task in the automotive industry. Thus, most of the research relies on short-term data logging or generating simulated data \cite{4}. Translating the customer service reports to practical dataset could offer a solution to feed data-driven fault detection algorithms.

Recently, due to the rapidly growing vehicles' technology, the conventional model/rule-based diagnostics methods are no longer efficient. Some studies have focused on a specific component developing diagnostics routines for maintenance procedures. Zhang et al. \cite{5} modeled a permanent magnet electric motor Using structural analysis to define a specific diagnostic system in electric vehicles. Meinguet et al. \cite{6} feed similar modeling methods with collected data from a test bench electrical traction system to introduce a DC-DC inverter diagnostic system. However, their research remained at a component level.

Data-Driven diagnostic approaches due to the system complexity have received more attention in recent years. Most of the researches have focused on a specific component using the methods that rely on limited proper sensors/controllers' logged data. Cheng et al. \cite{9} proposed a data-driven fault detection method, named the "Deep Slow Feature Analysis" method (DSFA), for the running gears of high-speed trains. The designed system used statistic algorithms to perform fault detection on the multi-dimensional running gear data in the test bench. The vehicle is equipped with six extra sensors to collect the data via timely and expensive test procedures. Prytz et al. \cite{7} applied supervised ML techniques to detect the failure of commercial trucks' air compressors. Logged on-board data collected over three years from a large number of trucks. Wolf et al. \cite{8} proposed an unsupervised learning data-driven diagnostics approach to detect faults by transferring the concept of deep embedded clustering for static data to multivariate in-vehicle time series. However, the collected data in both studies could not reach sufficient resolution to be applicable for accurate data mining algorithms.

Another approach, Interactive Voice Response (IVR), navigates customers through predefined options to address their issue, but with limited success \cite{10}. NLP is a computational technique that converts human language into smaller pieces, analyzes their relationships, and explores how they combine to form meaningful patterns \cite{11}. The NLP customer service techniques have been widely used to analyze call transcriptions, classify conversation topics, and identify customer sentiment. The NLP results could be a reliable source for companies to make solid decisions and improve customer service. Soysal \cite{12} has developed a clinical NLP graphical user interface toolkit that helps users to retrieve meaningful information from medical reports. In \cite{13}, the authors have accessed the radiology reports to create a rule-based algorithm to classify pulmonary oncology.

Yet, comprehensive approaches to apply the standard NLP methods in the vehicle diagnostic have not been explored in depth. Green \cite{14} introduced a pipeline for driver-vehicle voice interactions by concentrating on vehicle operator voice commands. In this way, Zheng \cite{15} developed an ML algorithm to enhance GPS routing commands for vehicle drivers. Therefore, introducing field-related preprocessing and feature extraction for unique NLP vehicle diagnostic applications is essential. In a similar study, Jalayer et al. \cite{20} classified hydroplaning crashes from free text police accident reports which include numerous domain-related expressions and abbreviations.

\subsection{Contribution}

The main contribution of this research is to propose an automated diagnostic model that evaluates vehicle failure claims and classifies them to different service departments. This is achieved by introducing domain-related NLP techniques and applying ML models on the free text vehicle service reports.  
The rest of the paper is organized as follows. Section II describes the dataset structures that is used to feed the models. 
Section III details preprocessing steps required to extract specific information from the customer service reports, and the feature selection process are explored.
In Section III, the Statistical and deep learning models are developed and evaluated to distinguish valid and vague service claims. Then, we compared classification methods used to diagnose failure groups.
The result of different models are discussed and compared in Section IV, and we conclude in Section V.

\section{DATASET}
\label{dataset}

This research used a dataset consisting of ten years of customer service calls to a fleet truck company. The given dataset was obtained from a private sector; additional details are omitted due to its proprietary nature. In addition, some of the names and labels are slightly modified for privacy protection. The sample data consists of 100,000 recorded service calls. Each call includes 11 main sections, as follows:
\begin{itemize}
\item "\textit{Service Department}" specifies one of the 16 possible departments. Each of them focuses on troubleshooting a specific vehicular component. The departments are shown in Table 1. Also, one department is titled "Vague" to represent unknown or unique service requests.  

\begin{table}[!ht]
\caption{Service Departments}
\label{table_example}
\begin{center}
\begin{tabular}{|c|c||c|c|}
\hline
1 & Controls& 2 & Vague\\
\hline
3 & Harness & 4 & Hydraulics\\
\hline
5 & PTO & 6 & Boom \\
\hline
7 & Maintenance & 8 & Test\\
\hline
9 & Rotation & 10 & Auger\\
\hline
11 & Outrigger & 12 & Digger \\
\hline
13 & Body & 14 & Chassis\\
\hline
15 & Electronics & 16 & Resale \\
\hline
\end{tabular}
\end{center}
\end{table}

\item "\textit{Service Call Log}"  contains a couple of words summary entered by the service representative summarizing the customer's complaints. It may be a combination of words, unit number or model, service date, or service number. It can also be blank if the representative could not summarize the service call into a sentence. In some cases, it can be general comments such as "service needed". Table 2 shows a couple of samples of Service Call Logs. 

\item "\textit{Service Detail}" consists of one paragraph to a page description of the service performed to date on the failed vehicle.  Table 2  shows examples of Service Details. A dash or a new line separates each service task. Dictation mistakes, connection words, and numbers are included in this section.
 Sentences in this section have better formatting compared to the Service Call Log section. This section usually does not follow a typical text structure. It may include the service date, vehicle model or vehicle identification number, parts replaced, and action performed to fix the issue. Technical abbreviations and expressions are often used in this section.

\item "\textit{Relation}" summarizes  whether the Service Call Log and the Service Detail have the same concept.  This section is manually annotated after service closes with one of the following:

\begin{enumerate}
  \item "\textit{False claim}" The information in Service Detail is different from Service Call Log. It could be due to inaccurate technician diagnostic or misleading information from the customer. 
  \item "\textit{Valid claim}" Service Call Log summarises the Service Detail adequately. This information is valuable and can get addressed to the proper service department.
  \item "\textit{Vague claim}" Service Call Log does not provide any information to route to the proper department. "Empty" Service Call Log or general terms like "unit failed" are included in this section
\end{enumerate}

\item "\textit{Age of unit}" 
Reflects the unit age at the incident. This is used for failure diagnostics of parts and components. Which would be useful information for system prognosis.
\item "\textit{Vehicle location}" 
contains the postal zip code of the vehicle service shop. This information reflects the geographic location vehicle typically used during operation.
\item "\textit{Customer company}" 
contains the company name associated with the vehicle since the vehicles are used commercially.
\item "\textit{Date of failure}" 
contains the date the vehicle failure was reported to the service center. In addition, it defines which day of the week and year it failed.
\item "\textit{Run-time}" 
reflects how long the vehicle has been in service since the service vehicles used are primarily stationary. 
\item "\textit{Ownership type}" 
 defines whether the vehicle was rented, leased, or purchased. 
\item "\textit{Time from last service}" 
describes how long the vehicle was used after its last service to determine if it was due for service at the time of failure.
\item "\textit{Phone operator identification}" 
 contains an identification of the phone operator or site technician who logged the request. This is useful as service reporting is subjective in nature. Both the Service Detail and Service Call Log are subject to the interpretation of the service provider. 
\end{itemize}

\begin{table}[h]
\caption{Service Call Log and Service Detail example}
\label{table_example_1}
\begin{center}
\begin{tabular}{|c|}
\hline
Service Call Log Example\\
\hline
SPACE007-Repairs from inspection\\
24506-replace winch rope unit is down\\
inoperable\\
3336-stuck in the air\\
\hline
 Service Detail Example \\
\hline
--cut off and replaced damaged area or repair--\\
--installed new hinges and pained area\\
replaced and adjusted transfer pin-- perform test\\
replaced gasket PN9700007824--had to cut off\\
inpected and cleaned area--cleaned up shaft\\
970000271 replaced, adjusted system pressure\\

Unit DWN--HYD inspected -- replaced related valve\\
\hline
\end{tabular}
\end{center}
\end{table}

\section{PROPOSED NLP-BASED PIPELINE}
\label{PROPOSED-NLP-BASED-PIPELINE}

\begin{figure}
  \centering
  \includegraphics[keepaspectratio, width=1.0\textwidth ]{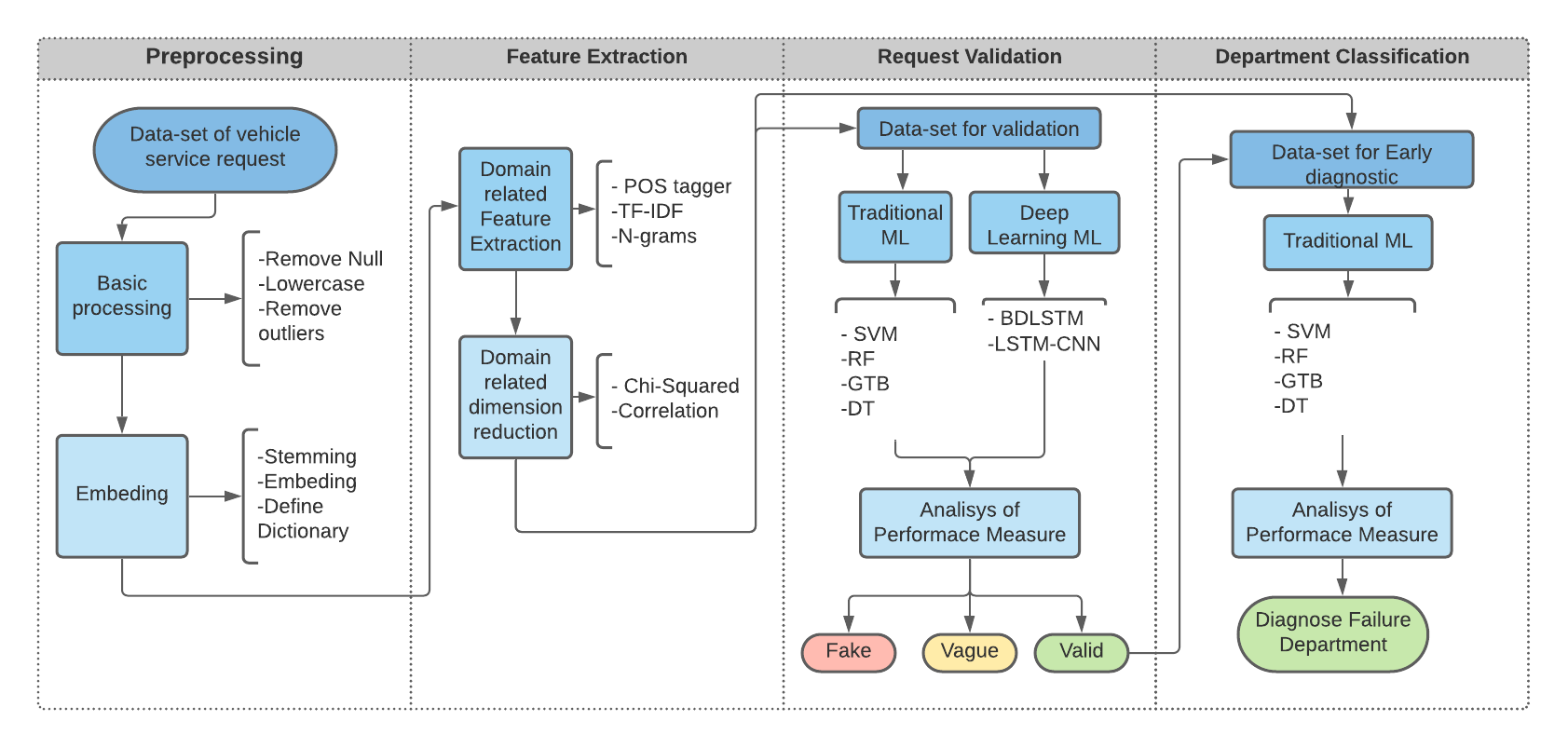}
  \caption{A simplified process flow-diagram implemented in this research. First, preprocessing techniques convert free text to meaningful information. Second, domain- related feature extraction defines the most reliable and efficient keywords and features used in the models. Third, the request validation section defines if the customer service claim is valid, fake, or vague. Finally, valid claims can be used for the last stage to perform department classification to route each request to one of the 16 different service departments.}
  \label{fig:3}
\end{figure}

A simplified process flow diagram implemented in this research is depicted in Figure \ref{fig:3}. 

\subsection{Pre-Processing}
Technical abbreviations and expressions are standard in both the Service Call Log and Service Detail sections. However, it is also evident that grammatical variances are present. As a result, preprocessing of data includes some domain-related and intuitive approaches detailed as follows.
We have defined extra lemmatization steps to enhance the preprocessing efficiency. Lemmatization is a process of reducing inflected words to their root and simple form. It is a process that maps related words to the same stemming root. As an example, the verb "break," "broken," and "breaking" from the dataset maps to the root word "break." In this specific example, we defined "brk" as break in lemmatization.  This step addresses the issue of adding new terms to the word lists when two words share the same information.
Words such as "an" in the majority of texts are defined as a definition word. Most algorithms recognize it as stop work and take it out from the test. However, in our application, it is the partial name of a product model. So any time "an" follows with a number, the algorithm won't consider it a stop word. Furthermore, some combinations of words such as "service needed"  are considered vague or irrelevant. They are removed from the data.

\subsection{Feature Extraction}
The feature extraction procedure is meant to extract a predefined set of features from the training text. Since machine learning algorithms cannot directly work on the raw text, vectors of weighted features are given to predictive models. The procedure of converting a raw set of selected words into a matrix (or vector) is also called “word embedding.” 
Like the preprocessing step, we have improved the efficiency of feature extraction by adding extra steps and techniques. 
There is no dictionary of staging terms to detect their presence in texts. This problem is well suited to regular expressions, at least for staging term recognition. We have introduced a specific library for this study using automotive expert’s support. The frequency of some of the words does not represent their importance in the text. For instance, the word “unit” or “vehicle” does not add any value to the algorithms and is removed from the feature list. So counting the work does not represent its importance in the text. Term Frequency Inverse Document Frequency (TF-IDF) is one of the standard feature extraction techniques in NLP. It reflects how important a word is in a document comparing the entire dataset. We have not observed relative strength in the word count used in a specific report compared to all of the dataset. Hence we did not peruse this approach in our study. Customer reports may contain technical abbreviations that are uncommon or unknown in traditional NLP tools.  In such instances, a Part of Speech Tagging (POS) may misclassify a word, leading to the unintended removal of essential terms. 
We modified the POS tagger and picked fine expressions to increase the precision to the point of utilization. POS tagger is a method to determine the type of terms used in the sentence \cite{19}. Further issues arise when using a tokenizer to parse the data. A tokenizer is a tool that separates a sentence into its component word tokens. This can be useful for extracting the words in a sentence that may then be passed into a POS tagger.
That said, this can be problematic when there is a combination of words that together represent one expression in the system. For example, “Upper valve” refers to a specific component in the vehicle.  A standard tokenizer, however, would split this phrase into two different words. This would lead to “upper” being classified as an adjective and “valve” being classified as a noun, resulting in the unintended removal of important information.
As another example, “unit down” in the report means the total failure of the vehicle. Again, a standard tokenizer would split this phrase into two words leading to “unit” being classified as a noun and “down” as an adjective. This illustrates the need to modify the tokenizer by incorporating domain-specific terminologies manually.
In this case, the most common nouns, adverbs, adjectives, and bigrams have been utilized to extract features from the report. N-grams are the continuous sequence of n-words in a sentence, considering it as a single unit \cite{19}.
Combinations of two words represent lots of components and services. As a result, bigram counts constitute an essential feature in this approach. For example, “PM Inspection” denotes a separate service department, and “Rotation Gearbox” represents a unique component in the vehicle.
A summary of the frequency of the different types of features in the dataset is presented in Table 3.
Table 4  shows the ten most frequent nouns and bigram features presented in the Service Detail section.   

\begin{table}[h]
\caption{Feature extraction summary}
\label{table_example_2}
\begin{center}
\begin{tabular}{|c||c|c|}
\hline
 & Service Call Log& Service Detail\\
\hline
Name & 75938 & 540930 \\
\hline
Verb & 3536 & 327446 \\
\hline
Adjective & 3434 & 53670 \\
\hline
Adverb & 245 & 110095 \\
\hline
Unigram & 24 & 228009 \\
\hline
Bigram & 31435 & 125075 \\
\hline
\end{tabular}
\end{center}
\end{table}

\begin{table}[h]
\caption{Most frequent nouns and bigrams in Service Detail}
\label{table_example_3}
\begin{center}
\begin{tabular}{|c|c||c|c|}
\hline
Noun & Count& Bigram & Count\\
\hline
Unit & 248470 & PM Inspection & 14850\\
\hline
Boom & 87380 & Dielectric Test & 13201 \\
\hline
Service & 43760 & Unit Function & 11890\\
\hline
Check & 41326 & Hydraulic Leak & 10459\\
\hline
Pole & 39375 & Boom Function & 8900 \\
\hline
Winch & 38541 & Upper Control & 7927\\
\hline
Inspection & 37044 & Annual PM & 7900 \\
\hline
Install & 35744 & Hydraulic Leak & 6566 \\
\hline
Hose & 34909 & Pole Guide & 6100 \\
\hline
Cylinder & 32700 & Rotation Gearbox & 4300 \\
\hline
\end{tabular}
\end{center}
\end{table}

In the last step of feature extraction, we have investigated different dimension reduction techniques for this application. 
 It is focused on reducing dimension of data so that the result data representation retains meaningful properties of the original data. Efficient dimension reduction would lead to decrease space required to store the data and less computation/training time.Furthermore, it resolves multicollinearity by removing redundant features.

\begin{figure}
  \centering
  \includegraphics[keepaspectratio, width=0.7\textwidth ]{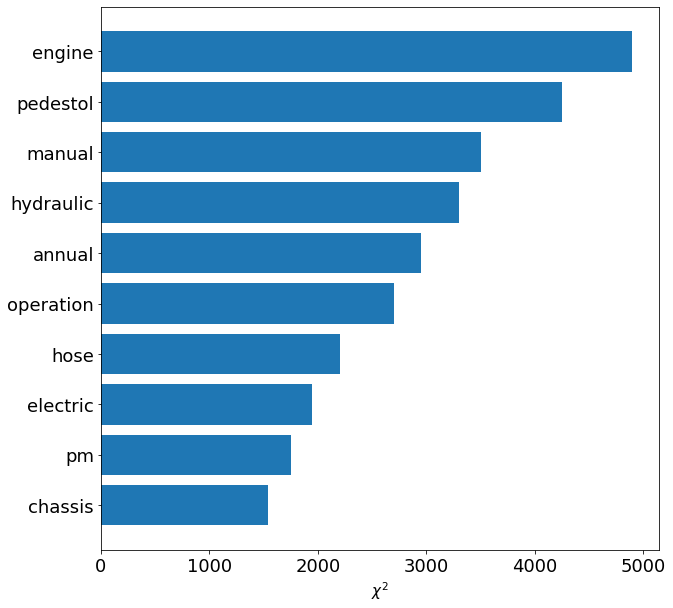}
  \caption{$\chi^2$ of the ten most significant features}
  \label{fig:4}
\end{figure}

Missing Value Ratio (MVR) is known as a primary dimension reduction method in data mining. MVR imputes the data or features that have a high ratio missing value. We have noticed that using this technique would cause losing meaningful information from the data despite standard text. For instance, "piston" has been used much less to other features in the whole dataset. The failure ratio of this part is rare. However, representation of this feature in the text has a higher weight typically in classification. We also used the Chi-Squared method to define the most efficient feature in the models and dimension reduction. Figure \ref{fig:4}  presents the distribution of the ten most significant features using the chi-square method.

\begin{figure}
  \centering
  \includegraphics[keepaspectratio, width=0.7\textwidth ]{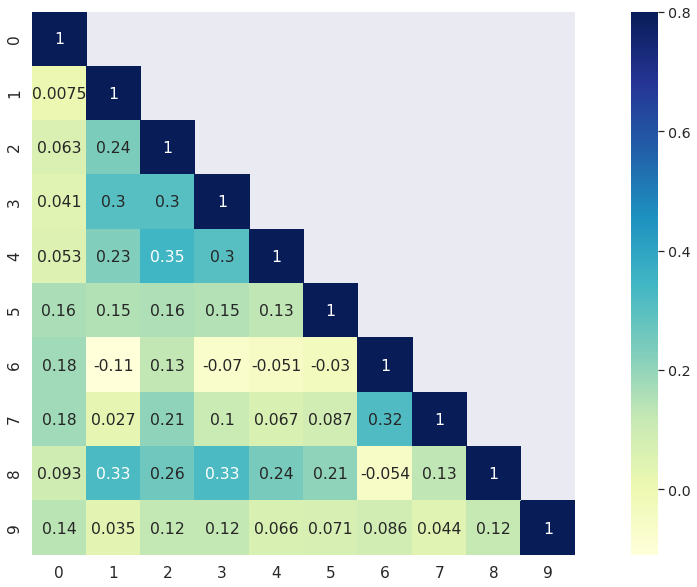}
  \caption{Feature correlation heatmap of the most ten important features}
  \label{fig:5}
\end{figure}

Finally, we used a heatmap to represent the correlation coefficient between the independent feature. It represents the strength of the linear relationship between two variables. In the machine learning context, the col-linearity between features can undermine the quality of a learning model. Usually, feature selection methods are deployed to reduce high dimensional feature sets to a smaller set for computational efficiency and reducing noise from redundant features. Figure \ref{fig:5}  shows the correlation between the features that are shown in Figure \ref{fig:4}.

\begin{figure}
  \centering
  \includegraphics[keepaspectratio, width=0.8\textwidth ]{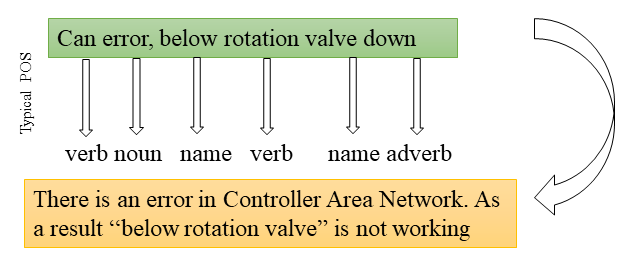}
  \caption{Example of how a common POS tagger recognise each sentence element and writer the indention}
  \label{fig:6}
\end{figure}

Figure \ref{fig:6}. represents a  sample sentence of the report.  It shows how a typical POS tagger defines each element of the sentence. Field-related techniques demonstrate in this study enabled us to be able to recognize the real operator intention. This specific example,” can,”  was defined in the dictionary, and a  noun term and  ”below rotation valve”  was picked as a single trigram feature.

\subsection{Request validation}

Request validation aims to use the attitude expressed towards a Service Call Log in the Service Detail section to determine whether or not a request is valid. This task is similar to stance analysis in other domains.  The outcome of stance analysis can be helpful for a variety of purposes. It has been widely used in the news to scale news accuracy and medical records to scale the diagnostic capability of the medical claim or reports. e.g., Chua  \cite{18} used this technique to prove that tweets stating facts were affirmed by 90\% of the tweets related to them, while tweets conveying false information were predominantly questioned or denied. Such a methodology can thus be used to remove vague or false service requests from the dataset. Using this approach, we removed service requests that the Service Call Log did not match the reports. As a result, the only service requests which the Service Call Log confirmed the Service Detail is used in the next section to classify the type of the request and route it to the related service department.  

Initially, we approached this problem using an SVM to perform a form of standard stance detection. We have carried out experiments with four different classifiers: SVM, Decision Tree, Gradient Tree Boosting, and Random Forest classifiers. Each classifier targets various features, including those based on features extracted in previous sections, unigrams, bigrams, models, and issue numbers. To evaluate the trained models, we used the ten-fold cross-validation technique. The initial results of the five mentioned classifiers were not satisfying and will be detailed in the results section. Therefore, we attempted to improve performance by producing a state-of-the-art classifier using CNN and BiLSTM networks.
Service Call Log and Service Detail are the input to the model, which are extracted and tokenized in words. Each word mapped to a vector representation, e.g., a word embedding, such that an entire report can be assigned to an \textit{s×d} sized matrix, where \textit{s} is the dimension of the embedding space and \textit{d} is the number of words in the report \textit{(\(d=500)\)}.
We applied zero padding such that all reports have the same  dimension \textit{\(x.r^{s'.d}\)}, where we chose \textit{\(s'=100\)}. We then applied several convolutions of different sizes to this matrix. A unique convolution involves a ﬁltering matrix \textit{\(w r^{h.d}\)} where \textit{h} is the size of the convolution, indicating the number of words it spans. The output \textit{\(c r^{s'.h+1}\)} is, hence, a concatenation of the convolution over all possible window of words in the report. In the paper, we used three ﬁlter sizes, and we used a total of 200 ﬁltering matrices for each ﬁlter size.

We then applied a max-pooling operation to each convolution. We used the max-pooling methodology to select the most important features for each convolution while discarding information about where in the service this feature is located. Using a CNN, we can efficiently extract the most practical combination of features in the embedding space, which is why we believe this model would be suitable for sentence classification. Finally, we used a  softmax layer to output the different class probabilities.
We added a dropout layer after the max-pooling layer and later a fully connected hidden neural network layer to reduce overfitting, with a dropout probability of 50\% during training. In the subsequent phase, the output of the CNN model is fed into the BiLSTM model.

We now describe the architecture of the LSTM system. The main architecture consists of two LSTM units. LSTM is a specific type of Recurrent Neural Network (RNN) built to work with sequential data by sharing their internal weights across the sequence. For each component in the series, that is, for each word in a sentence, an LSTM uses the standard word embedding and its previous hidden state to compute the next hidden state.

Typical LSTM unit does not consider post word information because the sentence is fed only in one direction to the model. It is a drawback of this algorithm in our application. To resolve this problem, we employ a Bidirectional LSTM, which is two LSTMs operating in opposite sequential orders. Finally, the outputs of both models are merged.

\begin{figure}
  \centering
  \includegraphics[keepaspectratio, width=1.1\textwidth ]{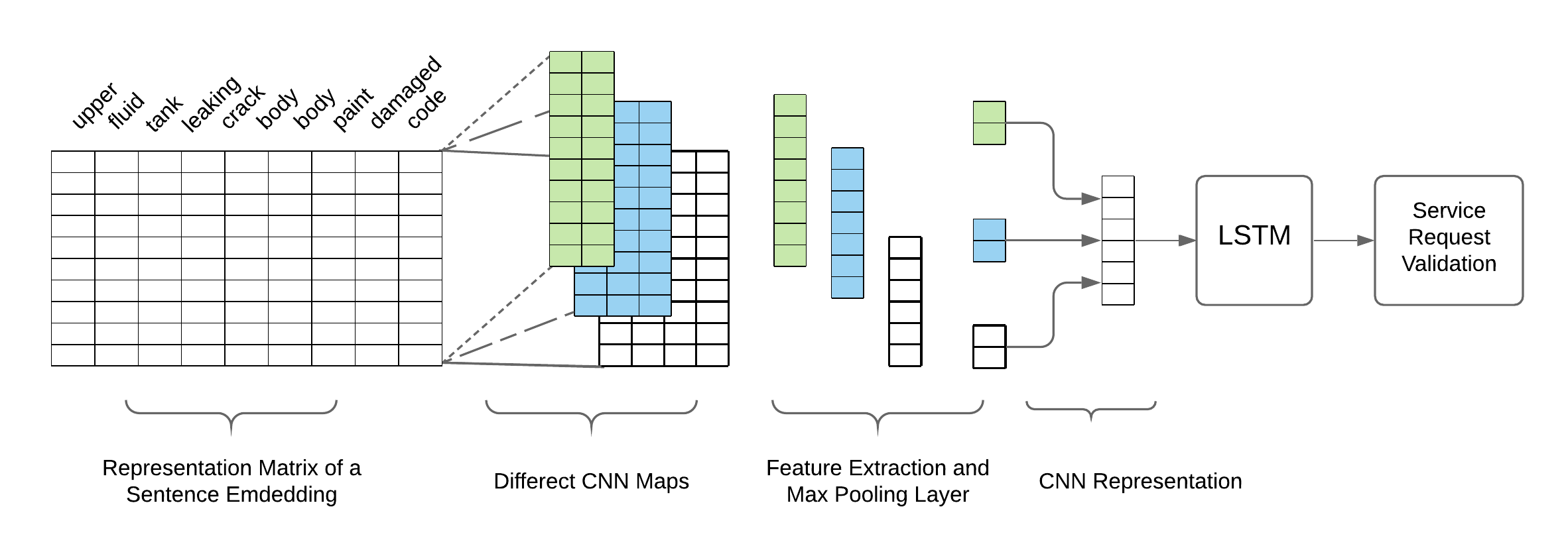}
  \caption{Combined CNN-BiLSTM Model architecture to determine accuracy of a service claim}
  \label{fig:7}
\end{figure}

Here again, we use dropout to reduce over-fitting; we add a dropout layer before and after the LSTMs and the fully connected hidden layer, with a dropout probability of 50 during training. Figure \ref{fig:7} demonstrates the overall structure of the CNN-BiLSTM architecture that has been used in this study.

\subsection{Classification Methods}

The next phase involves training classification models to route the requests to the 16 related service departments.
Complex and straightforward classification methods have been implemented on different types of texts. Since there have not been previous attempts to review customer service reports, we compare a set of well-known classifiers.  We trained four different classifiers including, SVM, Decision Tree, Gradient Tree Boosting, and Random Forest using the preprocessed data from selected feature lists. To evaluate the trained models, we used ten-fold cross-validation.

\section{RESULTS AND DISCUSSIONS}
\label{results_and_discussions}
To evaluate the model performance, different metrics such as Accuracy, F1-score, Recall, and Precision are used. \cite{20}. Proper evaluation method is closely tied to the nature of the problem and its expected outcomes.
In this research, weighted accuracy was used to evaluate the model performance.
  Weighted accuracy is a standard measure in this study since dataset is routed to multi-classes and equally distributed across different departments \cite{16}.

\[Accuracy=\sum_{k=1}^GW_i.\sum_{x:g(x)=k}I(g(x)=\hat{g(x)})\]

Here, \textit{G} is the number of classes, \textit{I} is the indicator function, which returns 1 if the classes match and 0 otherwise. The average Cost to address a service request varies across different departments. So our model is more sensitive to the performance of some individual classes. \textit{W} is the normalized weight assigned to an individual class which \(\sum_{k=1}^GW_i=1\). Table 5 summarize normalized weight used for each department in the studied dataset. As an example, based on historical service data, average Costs related to “boom” requests are departments are higher than “electrical” in the studied dataset.

\begin{table}[h]
\caption{Normalized weight associated to each department}
\label{table_example_5}
\begin{center}
\begin{tabular}{|c|c||c|c|}
\hline
Class & \(W_i\)& Class & \(W_i\)\\
\hline
Controls & 0.051& Vague & 0.082\\
\hline
Harness & 0.032 & Hydraulics & 0.034\\
\hline
PTO & 0.066 & Boom & 0.091 \\
\hline
Maintenance & 0.045 & Test & 0.072\\
\hline
Rotation & 0.066 & Auger & 0.070\\
\hline
Outrigger & 0.056 & Digger & 0.091 \\
\hline
Body & 0.073 & Chassis & 0.058\\
\hline
Electronics & 0.065 & Resale &  0.048\\
\hline
\end{tabular}
\end{center}
\end{table}

\subsection{Service request validation}

\begin{figure}
  \centering
  \includegraphics[keepaspectratio, width=0.63\textwidth ]{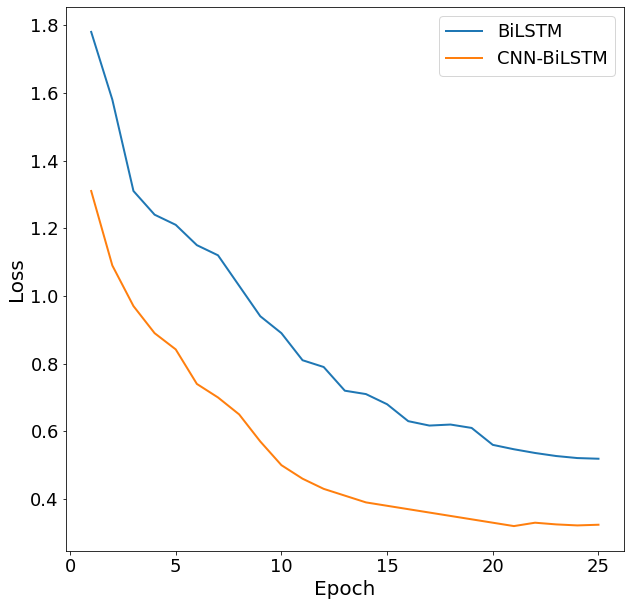}
  \caption{Loss in CNN-BiLSTM is  less then BiLSTM machine, it also decreases with the number of training epochs}
  \label{fig:8}
\end{figure}

The performance measures of the Service Validation section are compared in Table 6. The GTB achieved the highest performance in terms of accuracy, around 58\%. Meanwhile, RF achieved a relatively low and unsatisfactory performance. In the next step, we explored deep learning models as a means of improving performance. Since BiLSTM has achieved impressive results in other NLP applications such as media and medical, we developed BiLSTM algorithms on customer reports in this field. The BiLSTM used in the model is detailed in the previous section. This model achieved an initial accuracy of around 63\%, which is slightly improved over the other statistical models. However, it is not significant enough to justify using it due to the computational cost.  Looking at the report’s details, we noticed the different combination of the verbiage used in the text makes a significant impact in classifications. In the next step, we used a CNN to extract a meaningful combination of the features from the text before feeding it into the BiLSTM. The accuracy of this approach was around 85\%, which is a significant improvement over the previously tested models. The best performing result used a batch size of 200 after 15 epochs. Figure \ref{fig:8} shows that the loss function of the basic BiLSTM architecture for each epoch is improved using combined CNN-BiLSTM. 

\begin{table}[h]
\caption{Performance measures of different models for Service Request validation}
\label{table_example_6}
\begin{center}
\begin{tabular}{|c|c c c c c|}
\hline
Method & Accuracy & Sensitivity & Specificity & Precision & F-score\\
\hline
CNN-BiLSTM & \textbf{0.8476} & \textbf{0.8653} & \textbf{0.8313} & \textbf{0.9257} & \textbf{0.8450}\\

BiLSTM & 0.6343 & 0.5991 & 0.6754 & 0.6823 & 0.6379\\

GTB & 0.5843 & 0.6543 & 0.5185 & 0.5606 & 0.6038\\

RF & 0.4775 & 0.5029 & 0.4592 & 0.4001& 0.4456\\
\hline
\end{tabular}
\end{center}
\end{table}

\begin{figure}
  \centering
  \includegraphics[keepaspectratio, width=0.63\textwidth ]{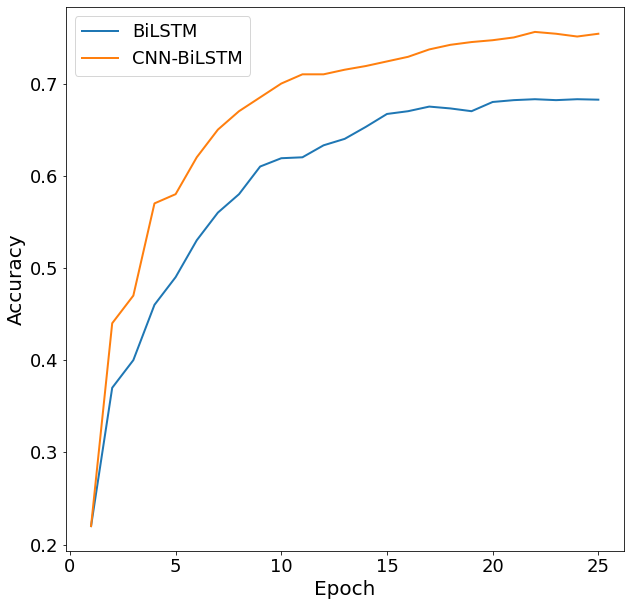}
  \caption{Accuracy compare in CNN-BiLSTM vs. BiLSTM machine, Accuracy improves with increasing epochs}
  \label{fig:9}
\end{figure}

Furthermore, loss consistently decreases with the number of training epochs. Figure \ref{fig:9} also compares the accuracy of the two methods. Similar to the loss function, the accuracy of the combined CNN-BiLSTM is approximately twice that of the basic BiLSTM model. For both models, accuracy improves with the number of training epochs.

\begin{table}[h]
\caption{Performance of different methods for classifying and routing Service Request to the relevant service department with and without domain-based NLP techniques at preprocessing and feature extraction stages along with CNN-BiLSTM based Request Validation}

\label{table_example_7}
\begin{center}
\begin{tabular}{|c c|c c c c c|}
\hline
 \scriptsize{\makecell{Service \\ Classification\\Model}}& \scriptsize{\makecell{CNN-BiLSTM \\Based\\ Request Validation}}& \scriptsize{Accuracy}& \scriptsize{Sensitivity}& \scriptsize{Specificity}& \scriptsize{Precision}& \scriptsize{F-score}\\

\hline
 GTB & No & 0.6018 & 0.5114 &0.7201 &0.7168 &0.5997\\
   & Yes &  \textbf{0.8585}  & \textbf{0.9080} &\textbf{0.8375} &\textbf{0.8137} &\textbf{0.8583}\\

RF & No & 0.6351 &  0.6270 &0.6433 &0.6395 &0.6327 \\
  & Yes &  0.8010  & 0.8548 &0.7541 &0.7659 &0.8079\\
DT & No & 0.5261& 0.5290 &0.5227 &0.5441 &0.6407\\
  & Yes &  0.6293  & 0.6672 &0.5921 &0.6162 &0.6407\\
SVM & No &  0.4301 & 0.3882 &0.4913 &0.5289 &0.4478 \\
  & Yes &  0.6121  & 0.5939 &0.6327 &0.6461 &0.6189\\
\hline
\end{tabular}
\end{center}
\end{table} 

\subsection{Classification and routing of service requests}

Service operators have around 70\% chance of predicting and routing the customer claim to the proper service department. Skilled operators, on average, have a 90\% chance of predicting and routing the customer claim to the proper service department. It is around 60\% for the new operator.
In our initial experiment, the 200 most frequent nouns, verbs, adjectives, and bigrams were chosen as input features. Classifiers achieved an average accuracy of around 70\% utilizing the most common feature. 
 Our results reveal that names and bigrams have the most impact on the accuracy of machines in this application. 
Comparing the accuracy results of different classifiers with various features, such as nouns, adverbs, and adjectives, revealed that words reflecting sentiments do, in general, impact classification performance. As a result, fewer adverbs and adjectives were chosen as features. 
 Due to the nature of the text, if a component or part name is mentioned, there is usually a failure associated with it. On the other hand, adjectives like “bad” or “failed” do not help classification accuracy.
We have observed the Random Forest accuracy is higher than the decision tree since it eliminates the variance in error observed with decision trees; however, it takes more time to train the model. Increased training time was not an issue in this study.
In general, the accuracy level we monitored in the application is slightly different than that of classification methods applied in other domains such as customer review or medical applications. This change is due to the nature of the text.
The results show by introducing the domain-specific preprocessing and feature extraction methods (explained in Section 3) in the first stage, the GTB classifier performance increased (25\% accuracy, 39\% sensitivity, 11\% specificity, 11\% precision, and 26\% f-Score). Performance of different classifications with and without domain-based NLP techniques at preprocessing and feature extraction stages along with CNN-BiLSTM based request validation are compared in Table 7.

\begin{figure}
  \centering
  \includegraphics[keepaspectratio, width=0.7\textwidth ]{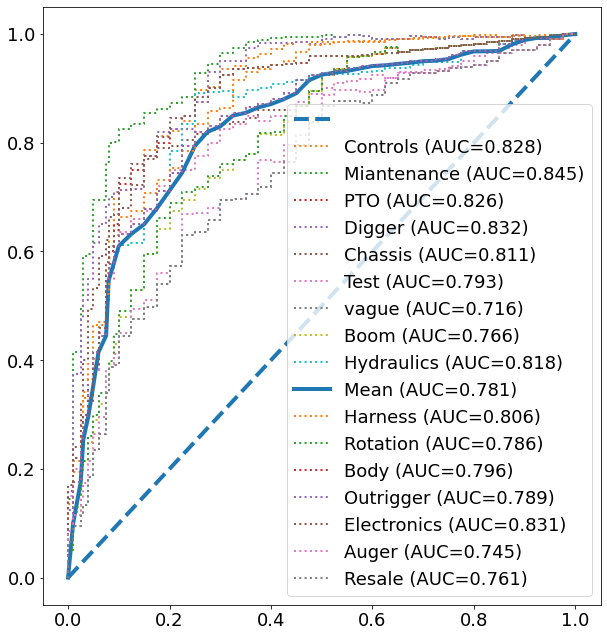}
  \caption{Random Forest classification results in terms of ROC and ROC-AUC for different Service Departments, respectively.}
  \label{fig:10}
\end{figure}

The performance is measured in terms of the average prediction accuracy in this study. We have also investigated the Area Under the Curve (AUC) between all 16 departments to decide. Usually, companies pay more attention to a specific type of failure considering other none technical decision factors such as cost to address the issue, severity of downtime of the vehicle. In that case, other mathematical calculations need to get considered to compare the classification methods. Figure \ref{fig:10} represents the AUC of ten service department classification using RF. The model has better success in some of the departments compared to the others. The AUC difference is due to the nature of vehicle failure. For example, high probability of engine failure if this word is identified in the text. On the other hand, “battery” does not have a good AUC. Usually, the majority of electrical component failures would cause battery failure.

Finally, the combined two-stage model for request validation and classification using CNN-BiLSTM/GTB shows ~8\% improvement in term of accuracy, compared to our previous work \cite{21}. The most relevant study to our work by Jalayer \cite{20} successfully applied RF to free text police accident reports to classify hydroplaning crashes with an accuracy of 0.6429, precision of 0.8136, and sensitivity of 0.6234. We do not have access to their code or dataset to make direct comparisons, but our model achieves a comparable accuracy level.

\section{CONCLUSIONS}
\label{conclude}

We developed domain-specific taxonomy, deployed specific preprocessing and feature extraction NLP techniques to extract significant information from free-text reports. Next, the efficiency of statistical and deep learning algorithms in validating customer service claims are compared. The most effective CNN-BiLSTM model could reach 0.84 accuracy and 0.92 precision in validating customer vehicle service requests.

Different classification methods are studied to route valid customer requests to the relevant service departments. We attained 85\% efficiency in term of accuracy, incorporating two-stage models of CNN-BiLSTM to filter fake/vague service claims and GTB to route the valid service claims to the admissible departments. As a result, CNN-BiLSTM/GTB pipeline exceeds 70\% success rate compared to the average operator. The outcome of this study would benefit future data-driven prognostic and diagnostic research in the vehicle industry.

\section{References}



\bibliographystyle{elsarticle-num} 


\nocite{*}
\bibliography{main.bib}
\end{document}